\documentclass{sig-alternate}

\usepackage{afterpage}
\usepackage{color}

\newcommand{\ignore}[1]{}

\begin{document}

\title{The Effects of Latency Penalties in Evaluating\\ Push Notification Systems}

\numberofauthors{4}
\author{
Luchen Tan, Jimmy Lin, Adam Roegiest, and Charles L. A. Clarke\\[1ex]
\affaddr{David R. Cheriton School of Computer Science}\\
\affaddr{University of Waterloo, Ontario, Canada}\\[1ex]
\affaddr{\{luchen.tan, jimmylin, aroegies\}@uwaterloo.ca, claclark@gmail.com}
}

\maketitle
\begin{abstract}
We examine the effects of different latency penalties in the
evaluation of push notification systems, as operationalized in the
TREC 2015 Microblog track evaluation. The purpose of this study is to inform
the design of metrics for the TREC 2016 Real-Time Summarization track,
which is largely modeled after the TREC 2015 evaluation design.
\end{abstract}

\section{Introduction}

There is emerging interest in building push notification systems that
filter social media streams such as Twitter to deliver relevant
content directly to users' mobile phones. The Microblog track at TREC
2015 operationalized the push notification task in the so-called
``scenario A'' variant of the real-time filtering
task~\cite{Lin_etal_TREC2015}. This evaluation forms the basis of the
Real-Time Summarization (RTS) track at TREC 2016. To help inform the
design of metrics for this new evaluation, in this short paper we examine the
effects of latency penalties by considering the impact of different
metric variants on runs submitted to TREC 2015. The primary goal of
this work is to provide a basis on which the evaluation metrics for
the RTS track at TREC 2016 can be developed.

\section{Background}

We assume that the reader is already familiar with the general setup
of the TREC 2015 Microblog track; otherwise, see Lin et
al.~\cite{Lin_etal_TREC2015} for details.
All experimental analyses in this paper are based on 
runs submitted to that evaluation.
Note the evaluation
consisted of two scenarios:\ ``scenario A'' push notifications and
``scenario B'' email digests. Here we focus only on push
notifications, as the latency penalty is not applicable for
scenario B.

At a high level, push notifications must relevant (i.e., on topic),
timely (i.e., the user desires news as soon as an event occurs), and
novel (i.e., the user does not want to see tweets that say basically
the same thing). Expected latency-dis\-counted gain (ELG), adapted
from the TREC Temporal Summarization track~\cite{Aslam_etal_TREC2014},
represents an attempt to capture these salient aspects. It is defined
as:
\begin{equation}
\frac{1}{N} \sum \textrm{G}(t)
\end{equation}
\noindent where $N$ is the number of tweets returned and
$\textrm{G}(t)$ is the gain of each tweet:\ non-relevant tweets receive
a gain of 0, relevant tweets receive a gain of 0.5, and highly-relevant
tweets receive a gain of 1.0.

A key aspect of this metric is its handling of redundancy and
timeliness:\ a system only receives credit for returning one tweet
from each cluster. Furthermore, a latency penalty is applied to all
tweets, computed as $\textrm{MAX}(0, (100 - d)/100)$, where the delay
$d$ is the time elapsed (in minutes, rounded down) between the tweet
creation time (i.e., when it was posted) and the putative time the tweet was pushed to the user. That is,
if the system delivers a relevant tweet within a minute of the tweet
being posted, the system receives full credit. Otherwise, credit
decays linearly such that after 100 minutes, the system receives no
credit even if the tweet was relevant.
Lacking any empirical guidance, the linear decay and the 100 minute
threshold represented arbitrary decisions made by the organizers.

Due to the setup of the task and the nature of interest profiles, it
is possible (and indeed observed empirically)\ that for some days, no
relevant tweets appear in the judgment pool. In terms of evaluation
metrics, a system should be rewarded for correctly identifying these
cases and not pushing non-relevant content. 
This is captured in the official metric as follows:\ If there are {\it no}
relevant tweets for a particular day and the system returns zero
tweets, it receives a score of one (i.e., perfect score) for that day;
otherwise, the system receives a score of zero for that day.
For the TREC 2015 topics, an empty run receives an ELG of 0.2471.
Recognizing that there is no relevant content to push appears to
be a difficult task, as most systems in TREC 2015 did not beat the
empty run in terms of ELG.

\afterpage{
\begin{figure}
\centering
\includegraphics[width=0.45\textwidth]{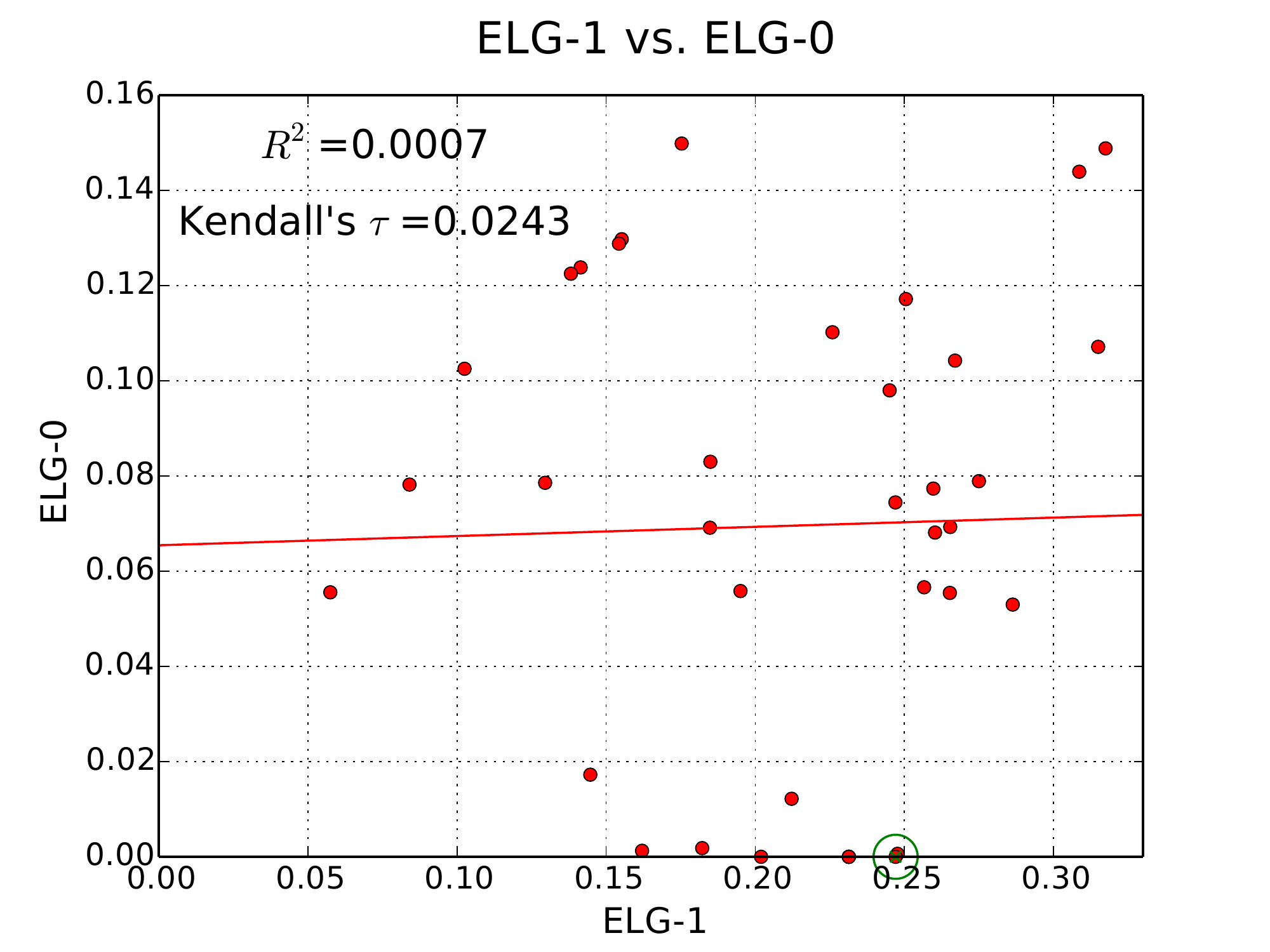}
\caption{ELG-1 vs. ELG-0 of all runs submitted to TREC 2015. Green
  circle indicates the empty run.}
\label{fig:elg_ncg}
\vspace{-0.2cm}
\end{figure}}

\begin{figure*}[t]
\centering\includegraphics[width=0.45\textwidth]{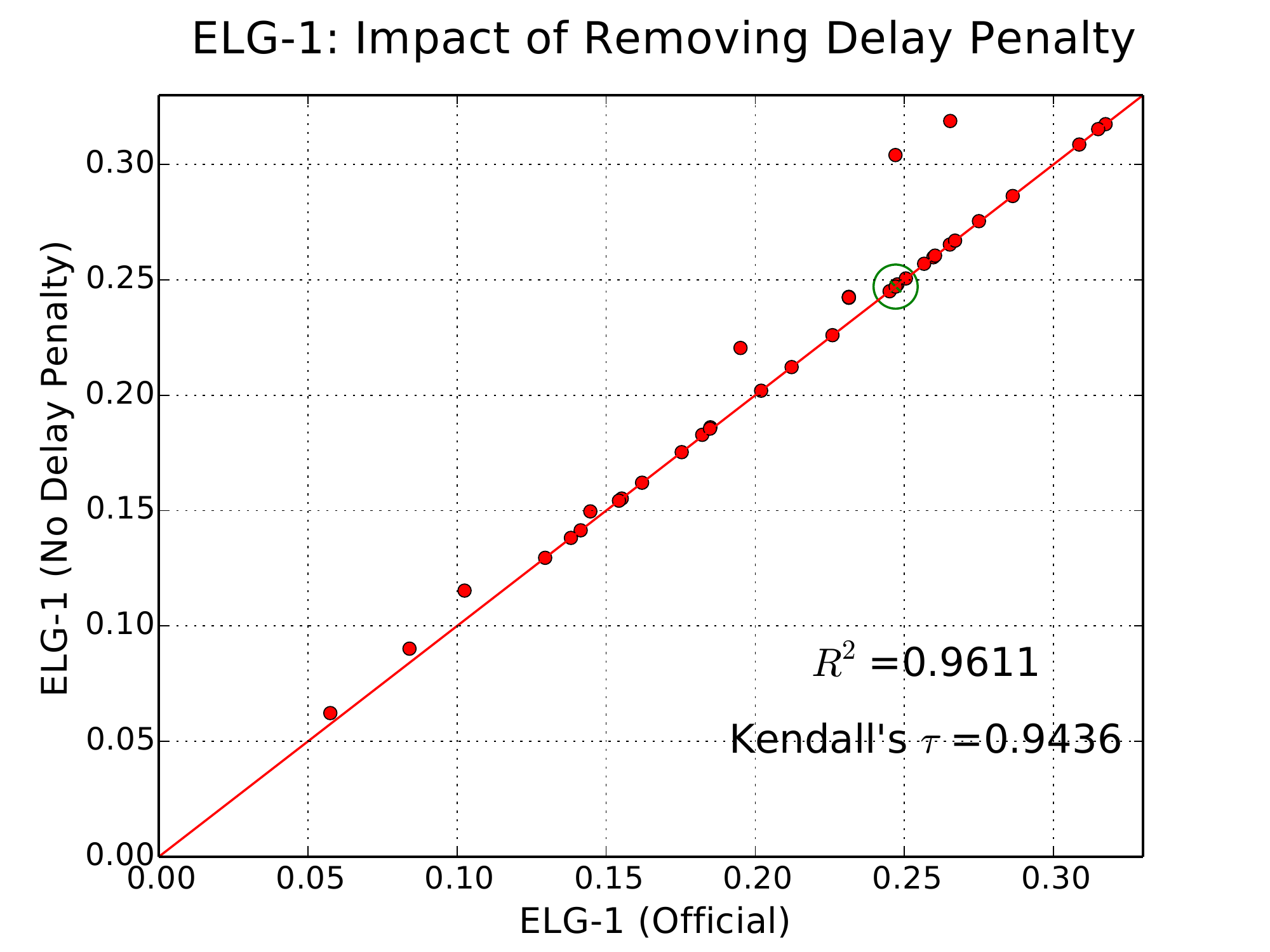}
\centering\includegraphics[width=0.45\textwidth]{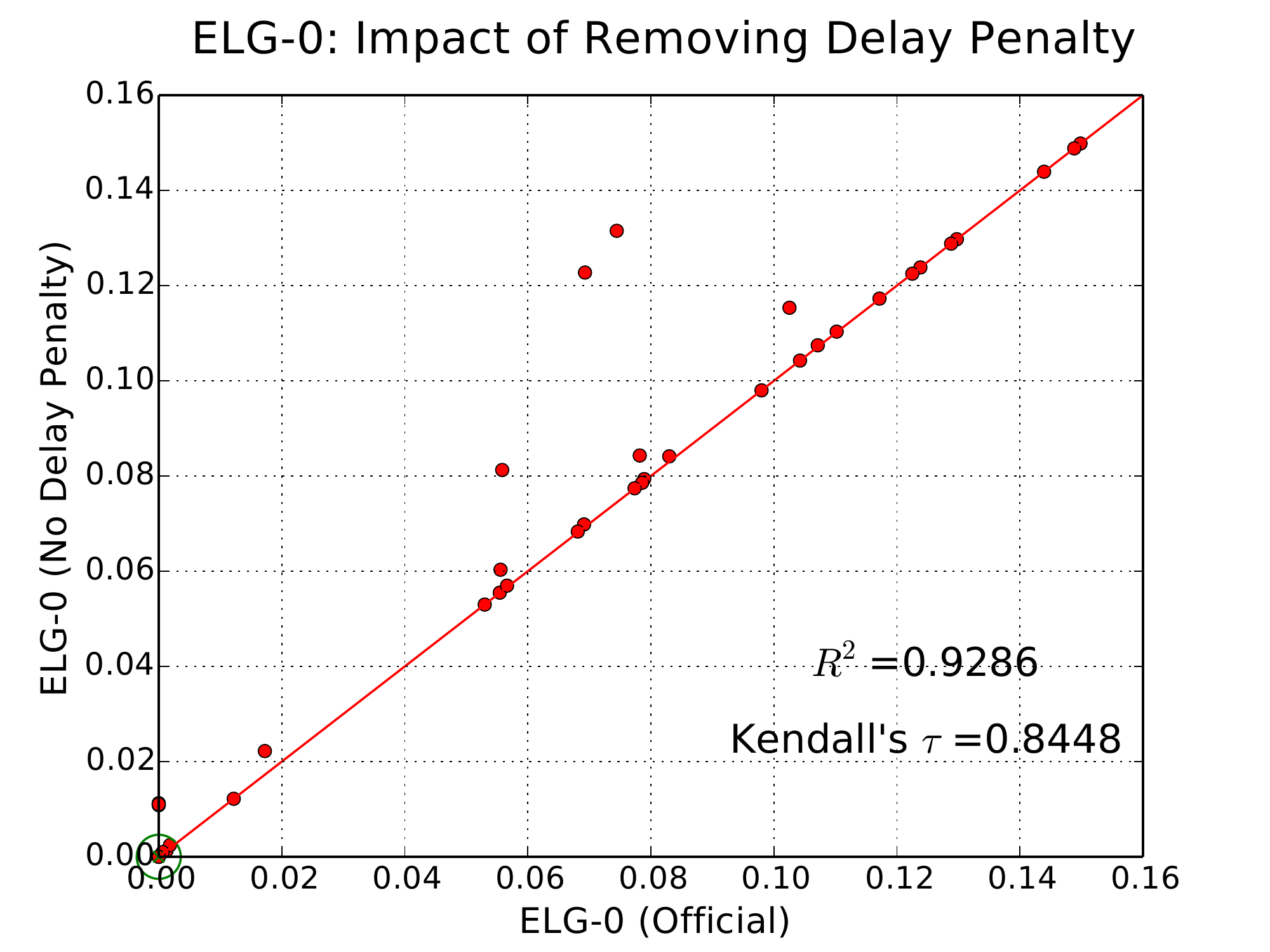}
\vspace{-0.2cm}
\caption{ELG-1 (left) and ELG-0 (right) of all runs submitted to TREC
  2015, comparing the official latency penalty definition with
  removing the latency penalty altogether. Green circles indicate empty runs.}
\label{fig:no_delay}
\vspace{-0.1cm}
\end{figure*}

Using the terminology of Tan et al.~\cite{Tan_etal_SIGIR2016a}, who
performed post hoc analyses of the TREC 2015 Microblog track evaluation, the official
ELG metric is called ELG-1. For rhetorical convenience, they call days
in which there are no relevant tweets for a particular topic (in the
pool) ``silent days''. On a silent day, according to ELG-1, the only two possible scores are one
(if the system remained silent) or zero (if the system pushed any
tweet). As an alternative, what if we did not reward systems for
remaining silent? That is, on a silent day, all systems receive a zero
score, no matter what they did. Tan et al.~called this variant metric
ELG-0, and here we adopt the same terminology.

Both variant metrics can be reasonably justified. ELG-1 makes sense
because we want to reward systems for knowing when to ``shut up'',
which is important if the user has many active interest profiles and
does not wish to be bombarded with notifications. On the other hand,
how would a user know to reward a system for staying silent---she has
no global knowledge of whether there actually were any relevant tweets
(in the evaluation, this global knowledge comes from pooling). Thus,
from an individual user's perspective, it makes sense just to give a
score of zero, regardless. Furthermore, ELG-1 introduces a sharp
discontinuity in the objective function, which makes system tuning difficult.

In Figure~\ref{fig:elg_ncg}, we show a scatterplot of ELG-1 vs.\ ELG-0
for all TREC 2015 runs. The green circle here and in all subsequent
scatterplots represents the empty run. 
The $R^2$ value reports the result of a linear regression, and
rank correlation is shown in terms of Kendall's $\tau$. 
We see no discernible relationship between the
two metrics, which shows that they are measuring different things.
Lacking user studies to provide empirical guidance, we have no way of determining which metric
better captures user preferences. For
these reasons, we report the results of subsequent analyses with respect to
both ELG-1 and ELG-0.

\section{Effects of the Delay Penalty}

The simplest way to quantify the impact of the latency penalty is to
remove it altogether. This analysis is shown in
Figure~\ref{fig:no_delay}, where for ELG-1 and ELG-0, we plot the
score of each run with and without the latency penalty. In both
scatterplots we show the diagonal $y=x$ (note, {\it not} the best fit
line) for reference. 
In this and all scatterplots, $R^2$ values report the results of
linear regressions, and rank correlations are shown in terms of Kendall's $\tau$.
As expected, all points lie above the diagonal,
since without the latency penalty system scores increase. We
were surprised, however, that the scores of many systems were exactly
the same, which meant that their algorithms made immediate decisions
with respect to each incoming tweet. In fact, there were only a few
outlier systems whose scores substantially changed, which meant that
they pushed tweets posted in the past. 

\begin{figure*}[p]
\centering\includegraphics[width=0.45\textwidth]{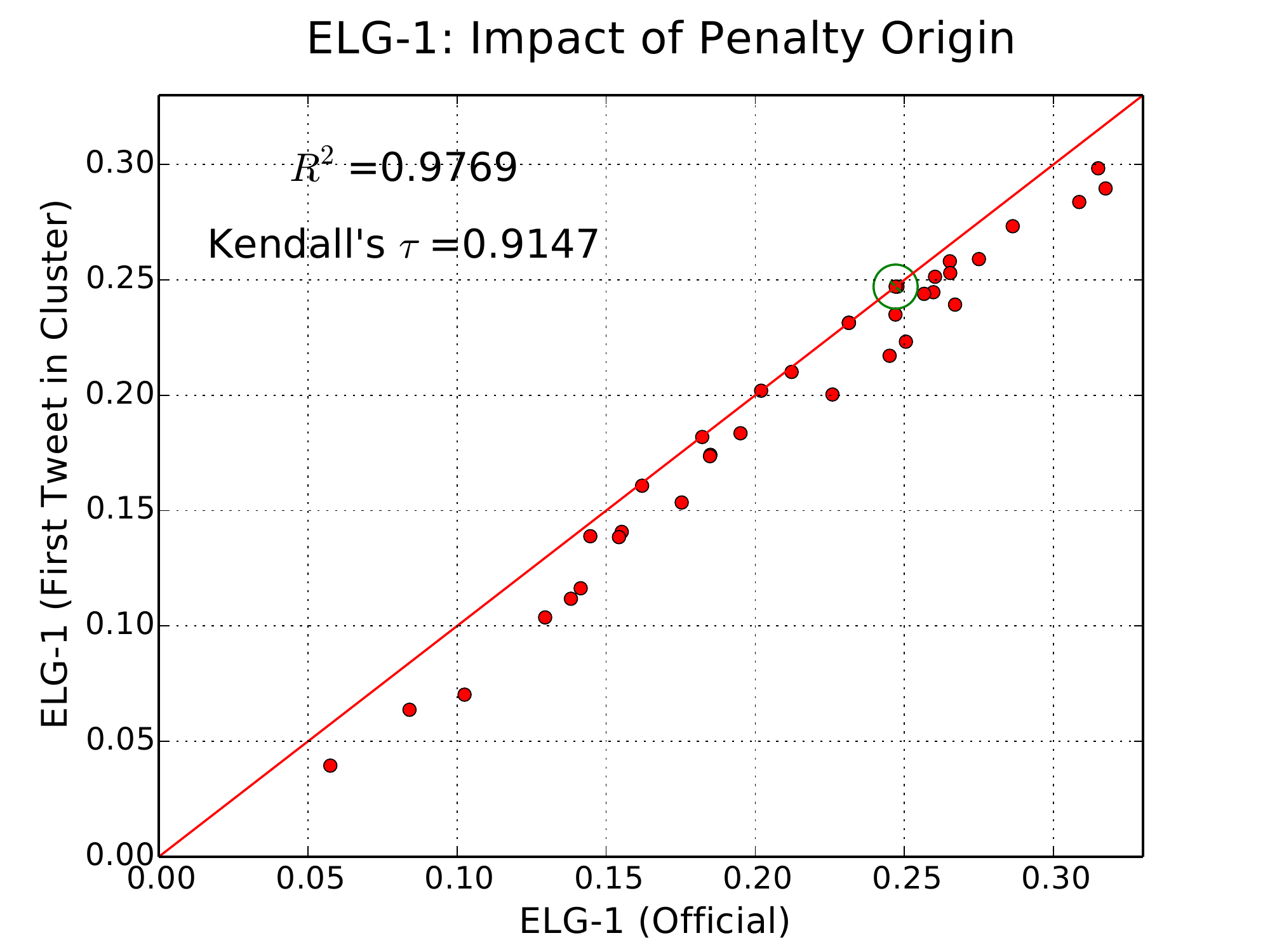}
\centering\includegraphics[width=0.45\textwidth]{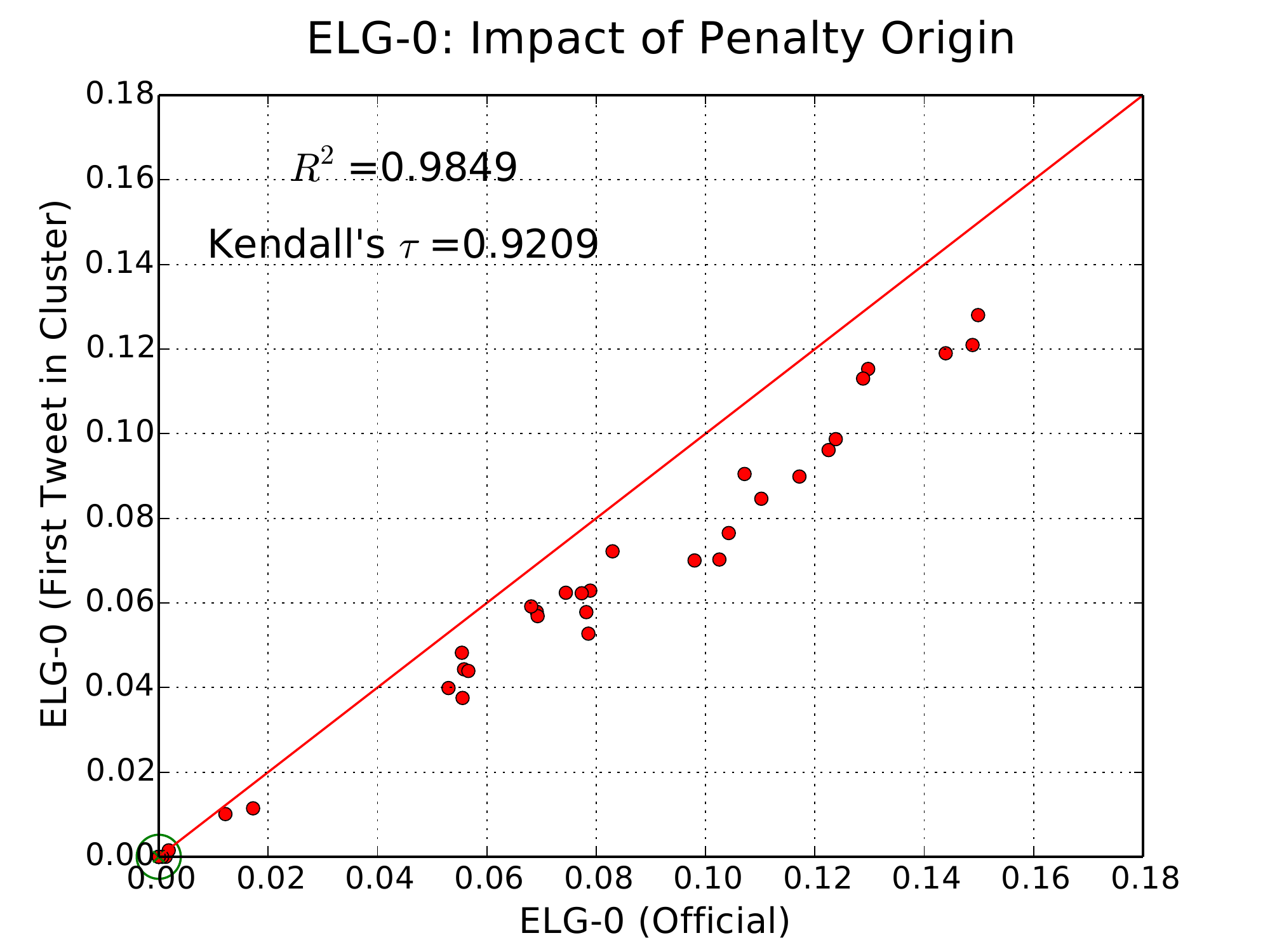}
\caption{ELG-1 (left) and ELG-0 (right) of all runs submitted to TREC
  2015, comparing the official latency penalty definition with
  computing the latency penalty with respect to the first tweet in
  each cluster. Green circles indicate empty runs.}
\label{fig:first_in_cluster}
\vspace{-0.2cm}
\end{figure*}

\begin{figure*}[p]
\centering\includegraphics[width=0.45\textwidth]{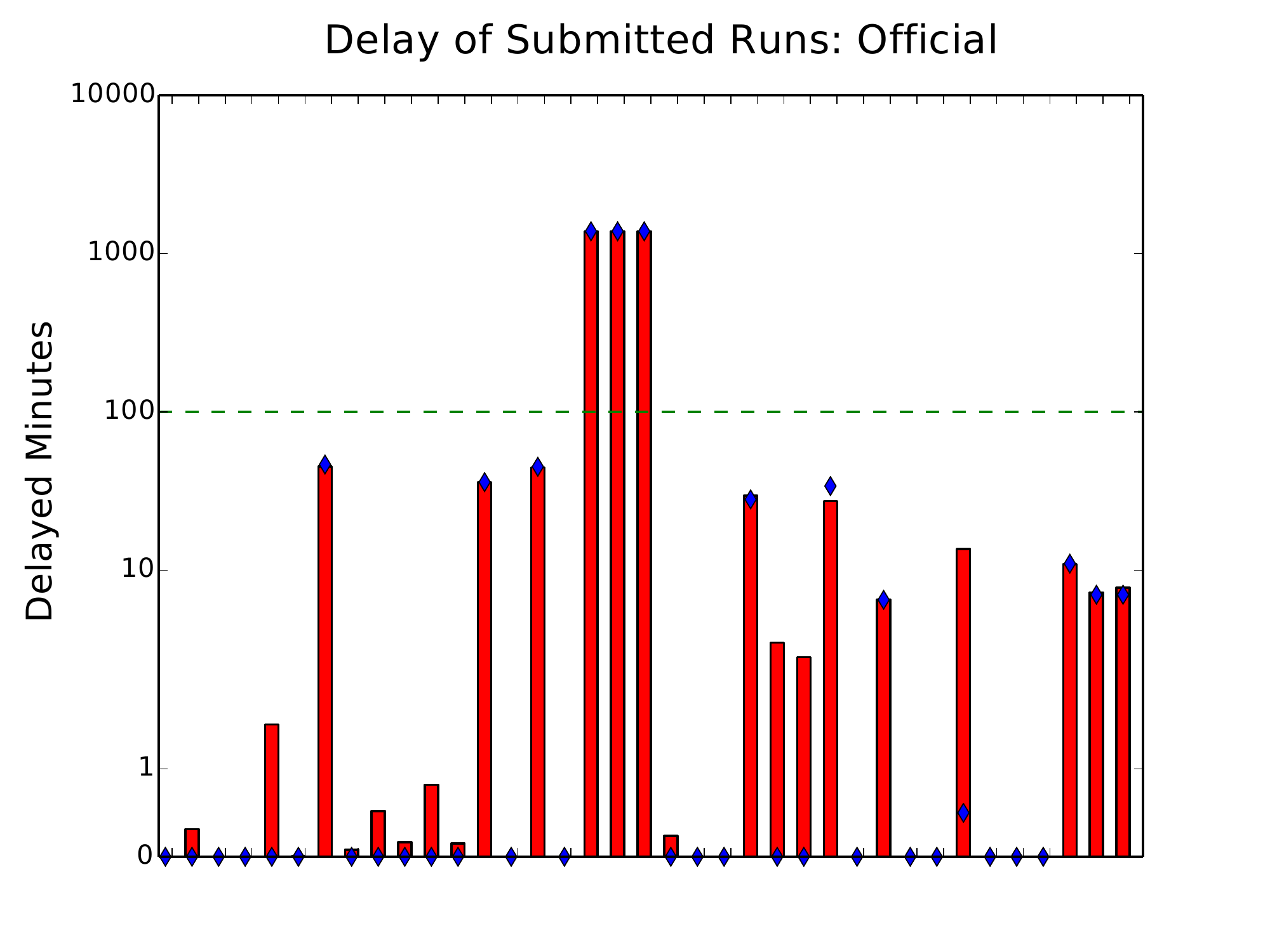}
\centering\includegraphics[width=0.45\textwidth]{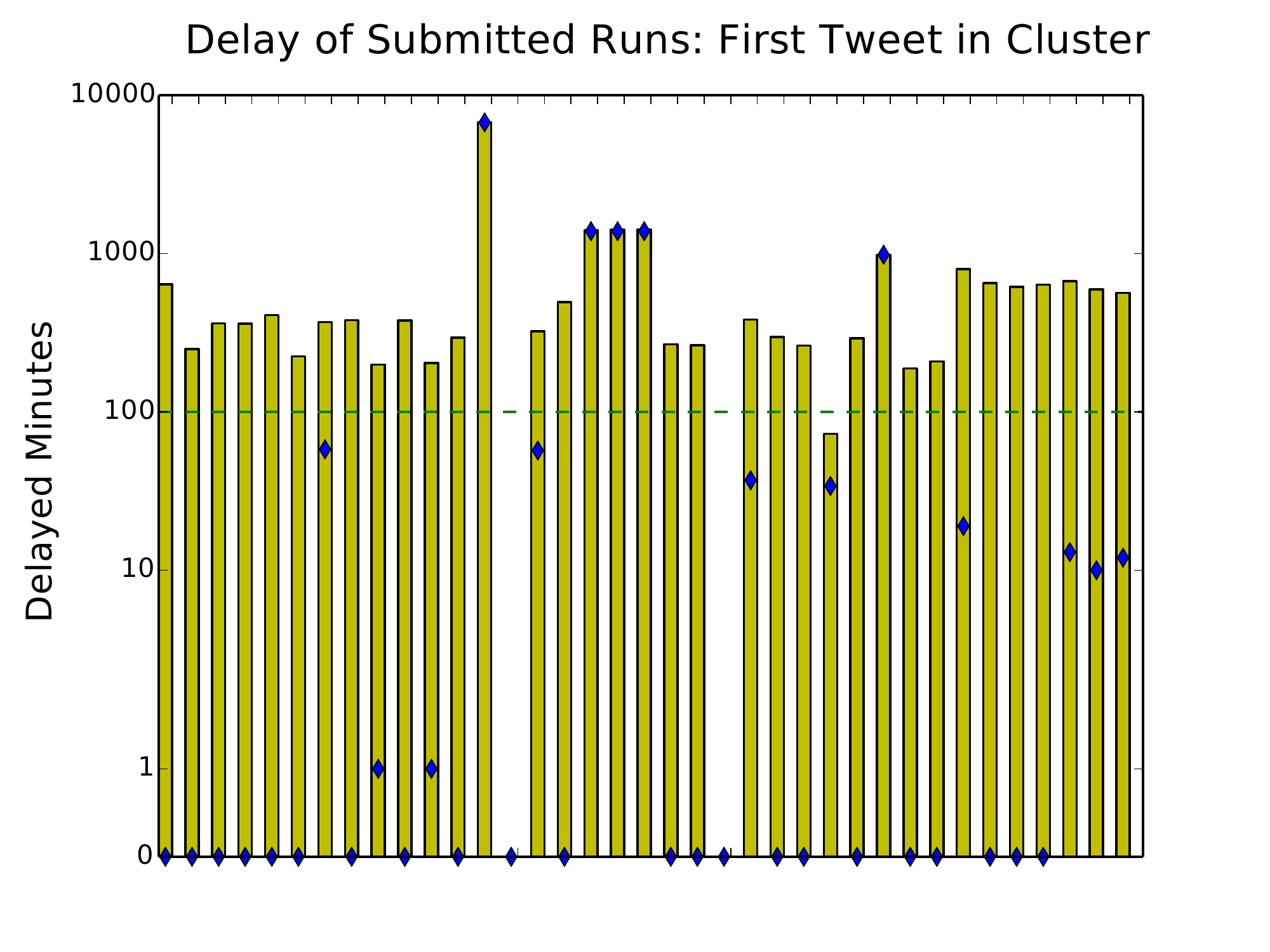}
\vspace{-0.5cm}
\caption{Quantifying the delay of each run in pushing tweets, with
  respect to the posted tweet (left) and with respect to the first
  tweet in each cluster (right). Runs are sorted in descending order
  of ELG-1.}
\label{fig:mean_delay}
\vspace{-0.2cm}
\end{figure*}

\begin{figure*}[p]
\centering
\includegraphics[width=0.45\textwidth]{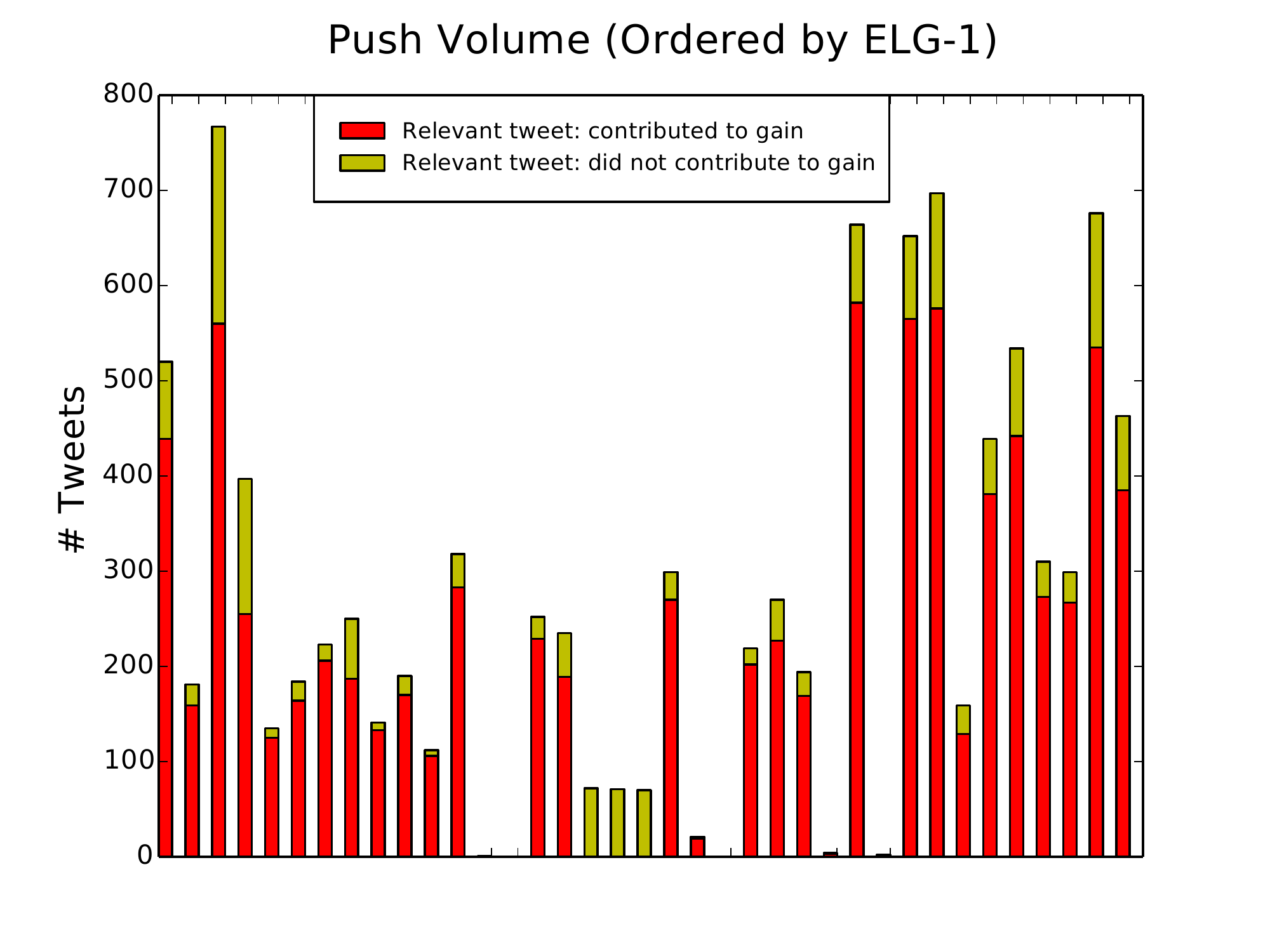}
\includegraphics[width=0.45\textwidth]{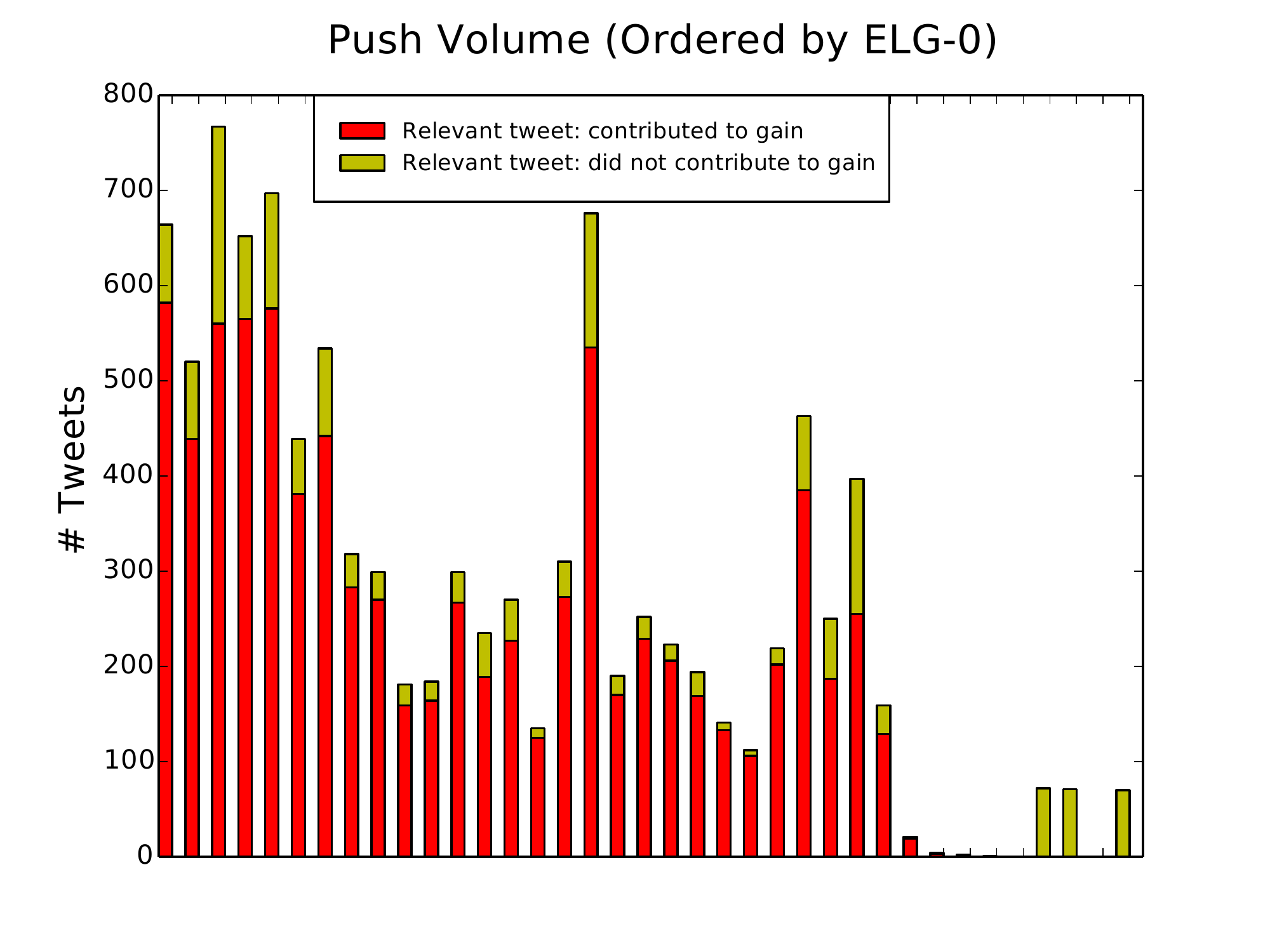}
\vspace{-0.5cm}
\caption{The push volume of each system, showing the number of
  relevant tweets pushed and the fraction that contributed to
  gain. The two bar charts differ in the sort order of the runs:\ on the
  left, in descending order of ELG-1, and on the right, in descending
  order of ELG-0.}
\label{fig:push_volume}
\vspace{-0.2cm}
\end{figure*}

We are quick to caution, however, that past system behavior is not
necessarily a good indication of system behavior in future
evaluations. In particular, TREC 2015 represented the first evaluation
of push notifications, and it is entirely possible that participants
focused on simple algorithms that did not attempt to model the
tradeoffs involved in pushing past tweets (i.e., accepting the latency
penalty for perhaps better relevance scoring).

Another noteworthy aspect of ELG (both the ELG-1 and ELG-0 variants) is that the
latency penalty is computed with respect to the pushed tweet, as
opposed to the first tweet in the cluster. Recall that in the
evaluation protocol, tweets are semantically clustered into
``equivalence sets'' that contain substantively the same
information. Let's consider the case where tweets $A$ and $B$ belong
to the same cluster, but tweet $B$ was posted three hours after tweet
$A$. Suppose system $P$ pushed tweet $A$ two hours after it was posted
and system $Q$ pushed tweet $B$ immediately when it was posted. Under
the official scoring metric, system $P$ would receive no credit
whereas system $Q$ would receive full credit; this doesn't make sense
since system $P$ conveyed the relevant information to the user
before system $Q$ did.

Recognizing this issue, it seems appropriate to compute the latency
penalty with respect to the first tweet in each cluster (which is
essentially what the Temporal Summarization track does). 
The effect of this change on system scores is shown in
Figure~\ref{fig:first_in_cluster}, where the scatterplots show each
run under the official score definition and the alternate computation
of the latency penalty with respect to the first tweet in each
cluster. In both scatterplots we show the diagonal $y=x$
for reference. As expected, all points lie below the diagonal since scores
decrease, but system rankings don't change much.

For another perspective, in Figure~\ref{fig:mean_delay} we show the
mean (bars) and median (diamonds) delay in pushing tweets by each
system, according to the official metric on the left, and with respect
to the first tweet in each cluster on the right. In these plots, we
only consider tweets that actually contributed to a run's score (i.e., yielded non-zero gain). Note
that the {\it y} axis is on a logarithmic scale in minutes. The bars
are arranged in descending ELG-1 score, from left to
right. From the bar chart on the left in Figure~\ref{fig:mean_delay}, we see that, indeed, most
systems always push immediately when a tweet is posted (if the
system thinks the tweet is relevant). We also see a few teams that
pushed tweets with a large delay---however, these are systems that
pushed very few results, and so their ELG-1 scores are fairly close to
that of the empty run.

One salient feature of the participating systems is that they vary
quite a bit in the volume of relevant tweets that they push. Because
of the reward associated with ``staying quiet'', systems can achieve
similar ELG scores with very different push volumes. This is shown in
Figure~\ref{fig:push_volume} (left), where each bar shows the total
number of relevant tweets that are pushed by
each system. The bars are arranged in decreasing ELG-1 score from
left to right. The red portions of the bars represent tweets that
contribute non-zero gain, while the tan portions of the bars represent
tweets that did not contribute any gain. These are either redundant
tweets or tweets pushed beyond the maximum acceptable latency (100 minutes)
to receive any credit.

We see that there are many cases where systems that
pushed more relevant tweets actually score lower than systems that
pushed fewer relevant tweets. Many of these are systems that always
push tweets no matter what---in other words, they don't know when to
``shut up''. In the range of middle-scoring runs, we see a number of
systems that barely push any content, and so their ELG-1 scores are
very close to that of the empty run (which, recall, was a
baseline that actually beats most systems). This effect is highlighted
in Figure~\ref{fig:push_volume} (right), where the runs are resorted
in terms of ELG-0 (but otherwise the bars are
exactly the same). Under this metric, systems are not rewarded for
staying quiet, and therefore systems that push more relevant tweets
tend to score higher.

As a final analysis, in Table~\ref{tab:clusters} we tally the number
of clusters for each topic, the number of singleton clusters (with
only a single relevant tweet), and singleton clusters expressed as a
percentage of all clusters. We see that most of the clusters are
singletons, which helps explain the results observed in
Figure~\ref{fig:mean_delay}:\ for singleton clusters, the latency
penalty is always computed with respect to the same tweet.

\begin{table}[t!]
\begin{center}
\begin{tabular}{r|r|r|r}
\hline
Profile & Clusters & Singletons & \% \\
\hline
\hline
MB228 & 3 & 3 & 100\%\\
MB236 & 97 & 40 & 41\%\\
MB242 & 73 & 47 & 64\%\\
MB243 & 119 & 70 & 59\%\\
MB246 & 171 & 127 & 74\%\\
MB253 & 3 & 3 & 100\%\\
MB254 & 32 & 27 & 84\%\\
MB255 & 15 & 12 & 80\%\\
MB260 & 1 & 0 & 0\%\\
MB262 & 67 & 48 & 72\%\\
MB265 & 58 & 44 & 76\%\\
MB267 & 39 & 29 & 74\%\\
MB278 & 35 & 30 & 86\%\\
MB284 & 51 & 42 & 82\%\\
MB287 & 71 & 58 & 82\%\\
MB298 & 8 & 7 & 88\%\\
MB305 & 3 & 2 & 67\%\\
MB324 & 3 & 3 & 100\%\\
MB326 & 20 & 10 & 50\%\\
MB331 & 25 & 8 & 32\%\\
MB339 & 10 & 9 & 90\%\\
MB344 & 818 & 594 & 73\%\\
MB348 & 37 & 22 & 59\%\\
MB353 & 10 & 9 & 90\%\\
MB354 & 18 & 3 & 17\%\\
MB357 & 7 & 7 & 100\%\\
MB359 & 8 & 7 & 88\%\\
MB362 & 55 & 49 & 89\%\\
MB366 & 97 & 79 & 81\%\\
MB371 & 103 & 64 & 62\%\\
MB377 & 12 & 9 & 75\%\\
MB379 & 28 & 15 & 54\%\\
MB383 & 16 & 13 & 81\%\\
MB384 & 14 & 11 & 79\%\\
MB389 & 17 & 14 & 82\%\\
MB391 & 75 & 61 & 81\%\\
MB392 & 17 & 6 & 35\%\\
MB400 & 87 & 81 & 93\%\\
MB401 & 311 & 236 & 76\%\\
MB405 & 5 & 5 & 100\%\\
MB409 & 9 & 5 & 56\%\\
MB416 & 8 & 6 & 75\%\\
MB419 & 56 & 36 & 64\%\\
MB432 & 44 & 41 & 93\%\\
MB434 & 276 & 257 & 93\%\\
MB439 & 67 & 63 & 94\%\\
MB448 & 1 & 0 & 0\%\\
\hline
Average & 66 & 49 & 74\%\\
\hline
\end{tabular}
\end{center}
\vspace{-0.2cm}
\caption{The total number of clusters and singleton clusters for each
  interest profile.}
\label{tab:clusters}
\vspace{-0.2cm}
\end{table}

\section{Conclusion}

Push notifications should be relevant, novel, and timely. The focus of
this work is the last property. Intuitively, systems should be
``punished'' for returning tweets late, hence the latency penalty
implemented in ELG. There is, however, little empirical
characterization of how real users would respond to push
notifications with increasing delay. Ultimately, user studies are
needed to ensure that metric definition and user needs actually align.

\bibliographystyle{abbrv}
\bibliography{latency} 

\end{document}